\documentclass[twocolumn,showpacs,preprintnumbers,amsmath,amssymb,pra]{revtex4}

\usepackage{color}

\usepackage{graphicx}
\usepackage{dcolumn}
\usepackage{bm}
\usepackage{epstopdf}
\usepackage{bbold}
\usepackage{amsthm}

\newtheorem{theorem}{Theorem}
\newtheorem{definition}{Definition}
\newtheorem{proposition}{Proposition}

\newcommand{\iden}{\mathbb{1}}
\newcommand{\tr}{\mathrm{tr}}

\newcommand{\mutinf}{I}
\newcommand{\clacorr}{J}

\newcommand{\ic}{\mathcal{I}}

\newcommand{\povm}{\mathrm{POVM}}

\newcommand{\im}{\mathrm{IM}}

\newcommand{\gio}{\mathrm{GIO}}

\begin{document}

\preprint{}

\title{Coherence non-activating measurement}

\author{Xueyuan Hu}
\email{xyhu@sdu.edu.cn}
\affiliation{School of Information Science and Engineering, Shandong University, Qingdao 266237, China}

\date{\today}

\begin{abstract}
We define the coherence non-activating measurement as the positive operator-valued measurement which gives the same result whether or not the coherence in a quantum state is destroyed. A connection is built between the coherence activating ability of a measurement and its ability to steer quantum states when coherence non-activating measurement is free. Then we study the quantum discord based on coherence non-activating measurement and its behavior under local incoherent operations. Our results contribute to the study of resource non-activating condition, which is a complementary to the well-studied resource non-generating condition.
\end{abstract}

\pacs{03.65.Ta, 03.65.Yz, 03.67.Mn}
\maketitle

\section{Introduction}
Quantum measurement, one of the fundamental elements in quantum theory, is an indispensable procedure in all the quantum information protocols. The quantumness of measurement can be revealed in the uncertainty principle \cite{PhysRevLett.60.1103,RevModPhys.89.015002}, the quantum steering \cite{PhysRevLett.113.160402,PhysRevLett.113.160403}, etc. In most of these regimes, two or more sets of measurement are involved, and the quantumness of measurement are viewed as the incompatibility of these sets of measurement \cite{PhysRevLett.122.130402,PhysRevLett.122.130403}. In the task of quantum steering, when a pure entangled state is shared, Alice can steer Bob's state if and only if she can implement sets of measurement which are not jointly measurable \cite{PhysRevLett.113.160402,PhysRevLett.113.160403}. It means that, the resource of steerability contained in quantum states can be activated only by measurements that are ``quantum''. Hence the quantumness of measurement can be viewed as its ability to activate the resource contained in quantum states.

In the resource theory of quantum coherence \cite{PhysRevLett.116.120404,RevModPhys.89.041003,HU20181}, a reference basis is fixed. States which are diagonal in the reference basis do not contain quantum coherence and are called incoherent states \cite{PhysRevLett.113.140401}. These states are considered free. The characterization of free operations attracts lots of interest, and different regimes are proposed, such as incoherent operations (IO) \cite{PhysRevLett.113.140401}, physically incoherent operations (PIO) \cite{PhysRevLett.117.030401}, strictly incoherent operations (SIO) \cite{PhysRevX.6.041028}, genuine incoherent operations (GIO) \cite{de_Vicente_2016}, etc., see Ref. \cite{PhysRevA.94.052336} for the comparison of them. All of these free operations are subset of maximal incoherent operations (MIO) \cite{arXiv:quant-ph/0612146,PhysRevA.94.012326}, which are defined as the whole set of quantum operations which do not generate coherence. Meanwhile, few attention has been paid to the non-activating condition \cite{PhysRevLett.118.060502}, which is a complementary to the non-generating condition.

Quantum discord \cite{RevModPhys.84.1655} is defined as the minimal discordance between the total correlation and the correlation that can be detected by local measurement. Here the local parties can implement the whole set of positive operator value measurement (POVM). Also, some works are focus on the discord defined on some specific measurement. For example, for measurement induced disturbance \cite{PhysRevA.77.022301} or diagonal discord \cite{ZW_DD}, the measurement is a projection to the eigenvector of marginal density matrix. When the measurement is a projection to a fixed basis, the corresponding discord is called the basis-dependent discord \cite{BD_discord,PhysRevX.6.041028}, which is closely related to quantum coherence.

In this article, we define the incoherent measurement as the POVM which can not activate the quantum coherence in any state. In other words, an incoherent measurement gives the same result whether or not the coherence in a quantum state is destroyed. We prove that the elements of an incoherent measurement are diagonal on the reference basis, and derive an inequality to witness whether a measurement can activate the coherence. Then we study the quantum discord based on incoherent measurement (QDI), which is similar to the traditional discord but the local measurement is limited to the set of incoherent measurement. Interestingly, the discord based on incoherent measurement equals to the basis dependent discord. The behavior of QDI under local incoherent operations are also explicitly studied.

\section{Coherence non-activating measurement}
A positive operator-valued measure (POVM) is associated with a set of positive operators $\{M_j\}$ satisfying $\sum_j M_j=\iden$. Instead of the measurement outcomes $j$, one cares about the probability distribution of the outcomes
\begin{equation}
p_j=\tr(\rho M_j),
\end{equation}
where $\rho$ is the state we put into the detection. If we get the same measurement result even through the coherence in $\rho$ is destroyed, we would say that the measurement $\{M_j\}$ coherence non-activating, namely, it is not able to detect the resource contained in $\rho$.

Let $\ic$ be the set of incoherent states, a coherence destroying map $\lambda$ satisfies two conditions \cite{PhysRevLett.118.060502}:\\
(1) $\lambda(\rho)\in \ic,\ \forall\rho$.\\
(2) $\lambda(\sigma)=\sigma,\ \forall\sigma\in\ic$.\\
A coherence non-activating measurement is then defined as follows.
\begin{definition}\label{Df:CNAM}
(Coherence non-activating measurement.) A quantum measurement $\{M_j\}$ is said to be coherence non-activating, if it satisfies
\begin{equation}
\tr(\rho M_j)=\tr(\lambda(\rho) M_j),\ \forall \rho,j.
\end{equation}
Here $\lambda$ is a coherence destroying map.
\end{definition}
By definition, the coherence destroying map is not unique \cite{PhysRevLett.118.060502}. However, if we require the coherence destroying map to be a completely-positive and trace-preserving (CPTP) map, then it is unique and given by the completely dephasing map $\Delta(\cdot)=\sum_j|j\rangle\langle j|(\cdot)|j\rangle\langle j|$ \cite{PhysRevA.95.062314}. In this case, a coherence non-activating measurement $\{M_j\}$ should satisfy
\begin{equation}
\tr(\rho M_j)=\tr(\Delta(\rho) M_j)=\tr(\rho\Delta(M_j)),\ \forall \rho,j.
\end{equation}
Hence each POVM element $M_j$ of a coherence non-activating measurement is diagonal in the incoherent basis. Because of this ``incoherent'' form, a coherence non-activating measurement is also called an incoherence measurement. We label the set of incoherence measurements as $\im$.

Apparently, all of the coherence non-activating measurements are jointly measurable, because they can be generated from incoherent projective measurement $\{|k\rangle\langle k|\}$ as $M_j=\sum_k m_{jk}|k\rangle\langle k|$. Conversely, if a measurement $S=\{S_j\}$ and the incoherent projective measurement $\{|k\rangle\langle k|\}$ are jointly measurable, then $S\in\im$. The reason is as follows. From the definition of jointly measurability \cite{PhysRevLett.113.160403}, $S$ and $\{|k\rangle\langle k|\}$ are jointly measurable, if and only if a measurement $G=\{G_\lambda\}$ exists such that both $S$ and  $\{|k\rangle\langle k|\}$ can be generated from $G$, i.e.,
\begin{eqnarray}
|k\rangle\langle k|&=& \sum_\lambda p(k|\lambda,0)G_\lambda,\label{eq:JM1}\\
S_j&=& \sum_\lambda p(j|\lambda,1)G_\lambda,\label{eq:JM2}
\end{eqnarray}
where $p(k|\lambda,0)$ and $p(k|\lambda,1)$ are conditional probabilities. Because $p(k|\lambda,0)$ are positive and each incoherent projector $|k\rangle\langle k|$ is of rank 1, Eq. (\ref{eq:JM1}) implies that every $G_\lambda$ is proportional to some incoherent projector. Substitute the form of  $G_\lambda$ to Eq. (\ref{eq:JM2}), we obtain that each $S_j$ is diagonal on the incoherent basis, so $S\in\im$. Remind that two sets of quantum measurement are not jointly measurable if and only if they can be used for quantum steering \cite{PhysRevLett.113.160403}, we arrive at the following proposition.

\begin{proposition}\label{prop:steer}
A measurement $M$ is not coherence non-activating, if and only if $M$ and the incoherent projective measurement $\{|k\rangle\langle k|\}$ can be used for quantum steering.
\end{proposition}

A consequence of this proposition is that, one can employ steering inequalities to judge whether a measurement is coherence non-activating or not. Suppose Alice wants to convince Bob that, besides incoherent measurement, she can also implement a measurement $M\notin\im$. From Proposition \ref{prop:steer}, if Alice can steer Bob's state, then Bob believes that Alice can indeed implement quantum measurement other than incoherent measurement. Inspired by the quantum steering inequality proposed in Ref. \cite{PhysRevLett.118.020402}, we derive an inequality for witnessing coherent measurement.

\begin{theorem}
Let $M=\{M_\alpha\}_{\alpha=0}^{n-1}$ be a quantum measurement on $d$-dimension systems. If an orthonormal basis $\{|\varphi_\alpha\rangle\}_{\alpha=0}^{d-1}$ exists such that the inequality
\begin{equation}
\sum_{\alpha=0}^{d-1}\langle\varphi_\alpha|M_\alpha|\varphi_\alpha\rangle>\sum_{i=0}^{d-1}\max_\alpha \left|\langle\varphi_\alpha|i\rangle\right|^2\label{eq:im_steering}
\end{equation}
holds, then $M\notin\im$.
\begin{proof}
Notice that one can always set $n\geq d$. If $n<d$, we can construct an equivalent measurement $M'$, with $M'_\alpha=M_\alpha$ for $\alpha<n$ and $M'_\alpha=0$ for $n\leq\alpha<d$.

If $M\in\im$, then we have $M_\alpha=\sum_{i=0}^{d-1}m_{\alpha i}|i\rangle\langle i|$. Because $n\geq d$, $M_\alpha\geq0$ and $\sum_{\alpha=0}^{n-1}M_\alpha=\iden$, we have $\sum_{\alpha=0}^{d-1}m_{\alpha i}\leq\sum_{\alpha=0}^{n-1}m_{\alpha i}=1$, and consequently,
\begin{eqnarray}
\sum_{\alpha=0}^{d-1}\langle\varphi_\alpha|M_\alpha|\varphi_\alpha\rangle &=& \sum_{i=0}^{d-1}\sum_{\alpha=0}^{d-1}m_{\alpha i}\langle\varphi_\alpha|i\rangle\langle i|\varphi_\alpha\rangle \nonumber\\
&=& \sum_{i=0}^{d-1}\sum_{\alpha=0}^{d-1}m_{\alpha i}\left|\langle\varphi_\alpha|i\rangle\right|^2 \nonumber\\
&\leq& \sum_{i=0}^{d-1}\max_\alpha \left|\langle\varphi_\alpha|i\rangle\right|^2.
\end{eqnarray}
Therefore, if the above inequality is violated for some orthonormal basis $\{|\varphi_\alpha\rangle\}_{\alpha=0}^{d-1}$, the quantum measurement $M$ is not an incoherent measurement.
\end{proof}
\end{theorem}

From the quantum steering point of view, Eq. (\ref{eq:im_steering}) means that Alice can steer Bob's state by implementing the measurement $M$ and the incoherent projective measurement $\{|i\rangle\langle i|\}$. Generally speaking, Eq. (\ref{eq:im_steering}) is not the necessary condition for quantum steering, so there are situations where Eq. (\ref{eq:im_steering}) is not satisfied for any $\{|\varphi_\alpha\rangle\}_{\alpha=0}^{d-1}$ even thorough $M$ is not incoherent.

However, for projective measurements with white noise, we prove that Eq. (\ref{eq:im_steering}) is the necessary and sufficient condition that the measurement is not incoherent. The measurement elements of a projective measurement with white noise $\tilde{\Pi}=\{\tilde{\Pi}_\alpha\}_{\alpha=0}^{d-1}$ can be written as
\begin{equation}
\tilde{\Pi}_\alpha=\lambda |\phi_\alpha\rangle\langle\phi_\alpha|+\frac{1-\lambda}{d}\iden,
\end{equation}
where $0\leq\lambda\leq1$ and $\{|\phi_\alpha\rangle\}$ are orthonormal basis which are not incoherent. Clearly, the measurement $\tilde{\Pi}$ is incoherent only when $\lambda=0$. If $\{|\phi_\alpha\rangle\}$ and $\{|i\rangle\}$ are mutually unbiased bases, we choose $|\varphi_\alpha\rangle=|\phi_\alpha\rangle,\forall\alpha$, and Eq. (\ref{eq:im_steering}) becomes $d\lambda+(1-\lambda)>1$, which holds for $\lambda\neq0$. If $\{|\phi_\alpha\rangle\}$ and $\{|i\rangle\}$ are not mutually unbiased, we choose $\{|\varphi_\alpha\rangle\}$ to be mutually unbiased with $\{|i\rangle\}$ but not with $\{|\phi_\alpha\rangle\}$. Hence the right-hand side of Eq. (\ref{eq:im_steering}) equals 1, and the left-hand-side reads
\begin{equation}
\sum_{\alpha=0}^{d-1}\langle\varphi_\alpha|M_\alpha|\varphi_\alpha\rangle = 1+\lambda\left[\sum_{\alpha=0}^{d-1}\left|\langle\phi_\alpha|\varphi_\alpha\rangle\right|^2-1\right].
\end{equation}
Because $\{|\varphi_\alpha\rangle\}$ and $\{|\phi_\alpha\rangle\}$ are not mutually unbiased, we can arrange the ordering of $\{|\varphi_\alpha\rangle\}$ such that $\sum_{\alpha=0}^{d-1}\left|\langle\phi_\alpha|\varphi_\alpha\rangle\right|^2>1$. Hence the left-hand-side is strictly larger than 1 if $\lambda\neq0$. This completes the proof.


\section{Quantum discord based on coherence non-activating measurements}
Before study quantum discord based on coherece non-activating measurements, we briefly review the definition of traditional quantum discord. For a bipartite state $\rho_{AB}$, the total correlation between $A$ and $B$ is quantified by the mutual information $\mutinf_{A:B}(\rho_{AB})=S(\rho_A)+S(\rho_B)-S(\rho_{AB})$, where $\rho_{A(B)}=\tr_{B(A)}(\rho_{AB})$ is the reduced density matrix of system $A(B)$, and $S(\rho)=-\tr(\rho\log_2\rho)$ is the von Neumann entropy. The maximal amount of information that can be revealed by local POVM on $A$ is called the classical correlation $\clacorr_{B|A}(\rho_{AB})=\max_{\{M_\mu\}\in\povm}[S(\rho_B)-\sum_\mu p^{M_\mu} S(\rho_{B|M_\mu})]$, where $p^{M_\mu}=\tr(\rho_{AB}M_\mu^A\otimes \iden^B)$ is the probability to get the measurement result $\mu$ and $\rho_{B|M_\mu}=\tr_A(M_\mu^A\otimes \iden^B)/p^{M_\mu}$ is the resulted state of $B$ after the measurement. The difference between total correlation and classical correlation is called quantum discord $\delta_{B|A}(\rho_{AB})=I_{A:B}(\rho_{AB})-J_{B|A}(\rho_{AB})$.

Now we are ready to define the incoherent correlation and quantum discord based on incoherent measurement.
\begin{definition}\label{def:qdi}
For a bipartite state $\rho_{AB}$, the incoherent correlation on $A$ is defined as the maximal information gain about $B$ as a result of an incoherent measurement on $A$
\begin{eqnarray}
J^\mathrm{I}_{B|A}(\rho_{AB}):=\max_{\{M_\mu\}\in\im}{\left[S(\rho_B)-\sum_\mu p^{M_\mu} S(\rho_{B|M_\mu})\right]},
\end{eqnarray}
where $p^{M_\mu}=\tr(M_\mu\rho_{AB})$ and $\rho_{B|M_\mu}=\tr_A(M_\mu\rho_{AB})/p^{M_\mu}$ are the probability and the resulted state of $B$ for the measurement result $\mu$. The quantum discord based on the incoherent measurement (QDI) is defined as the difference between the mutual information and the incoherent correlation
\begin{eqnarray}
D^\mathrm{I}_{B|A}(\rho_{AB}):=\mutinf_{A:B}(\rho_{AB})-J^\mathrm{I}_{B|A}(\rho_{AB}).\label{eq:qdi}
\end{eqnarray}
\end{definition}

In the definition of traditional quantum discord $\delta_{B|A}(\rho_{AB})$, the optimization is taken over the whole set of POVM, and the optimal measurement need not to be projective. Here, for the discord based on incoherent measurement $D^\mathrm{I}_{B|A}(\rho_{AB})$, the optimization is restricted to the set of incoherent measurement. Therefore, $D^\mathrm{I}_{B|A}(\rho_{AB})$ is lower bounded by $\delta_{B|A}(\rho_{AB})$, which is nonnegative. In the following we prove that the optimal incoherent measurement which reaches the minimization in $D^\mathrm{I}_{B|A}(\rho_{AB})$ is just the projection to incoherent basis.

\begin{theorem}\label{th:qdi}
The discord based on incoherent measurement has the following equivalent expressions
\begin{eqnarray}
\label{eq:qdi1}D^\mathrm{I}_{B|A}(\rho_{AB})&=&\sum_{i=0}^{d-1}p_iS\left(\rho_B^i\right)+S(\rho_A)-S(\rho_{AB})\\
\label{eq:qdi2}D^\mathrm{I}_{B|A}(\rho_{AB})&=&\mutinf_{A:B}(\rho_{AB})-\mutinf_{A:B}(\rho_{\tilde AB}),\\
\label{eq:qdi3}D^\mathrm{I}_{B|A}(\rho_{AB})&=&C_r(\rho_{AB})-C_r(\rho_{\tilde AB})-C_r(\rho_A).
\end{eqnarray}
Here $p_i=\tr\left[\left(|i\rangle_A\langle i|\otimes\iden_B\right)\rho_{AB}\right]$ and $\rho_{B|i}=\tr_A\left[\left(|i\rangle_A\langle i|\otimes\iden_B\right)\rho_{AB}\right]/p_i$ are the probability and the resulted state of $B$ after Alice implement incoherent projective measurement and get the result $i$, $\rho_{\tilde AB}=\Delta_A\otimes\iden_B(\rho_{AB})$, and $C_r(\rho)=S(\Delta(\rho))-S(\rho)$ is the relative entropy of coherence.
\begin{proof}
We first prove the equivalence between Eqs. (\ref{eq:qdi}) and (\ref{eq:qdi1}). Because the measurement $M$ on $A$ is coherence non-activating, the measurement element is diagonal in the incoherence basis $M_\mu=\sum_{i=0}^{d-1}m_{\mu i}|i\rangle\langle i|$, and the resulted state of $B$ for the measurement result $\mu$ is then written as
\begin{eqnarray}
\rho_{B|M_{\mu}}&=&\frac{1}{p^{M_\mu}}\tr_A\left[\left(\sum_{i=0}^{d-1} m_{\mu i}|i\rangle_A\langle i|\otimes\iden_B\right)\rho_{AB}\right]\nonumber\\
&=&\sum_{i=0}^{d-1}\frac{m_{\mu i} p_i}{p^{M_\mu}}\rho_{B|i},
\end{eqnarray}
Notice $\tr(\rho_{B|M_\mu})=\tr(\rho_{B|i})=1$, we have $\sum_{i=0}^{d-1}\frac{m_{\mu i} p_i}{p^{M_\mu}}=1$, and then $\{\frac{m_{\mu i} p_i}{p^{M_\mu}}\}_i$ is a probability distribution for all $\mu$. By the convexity of Von Neumann entropy, $S(\rho_{B|M_\mu})\geq \sum_{i=0}^{d-1}\frac{m_{\mu i} p_i}{p^{M_\mu}}S(\rho_{B|i})$. Hence, the following inequality holds for all incoherent measurement $\{M_\mu\}$:
\begin{eqnarray}
\sum_\mu p^{M_\mu} S(\rho_{B|M_\mu})\geq\sum_{i=0}^{d-1}p_iS\left(\rho_{B|i}\right).
\end{eqnarray}
On the other hand, $\min_{\{M_\mu\}\in\im}{\sum_\mu p^{M_\mu} S(\rho_{B|M_\mu})}\leq\sum_{i=0}^{d-1}p_iS\left(\rho_B^i\right)$, because the incoherent projective measurement belongs to $\im$. Therefore, Eqs. (\ref{eq:qdi}) and (\ref{eq:qdi1}) are equivalent.

Eq. (\ref{eq:qdi2}) is equivalent to Eq. (\ref{eq:qdi1}) because $\rho_{\tilde A B}=\sum_{i=0}^{d-1}p_i|i\rangle\langle i|\otimes\rho_B^i$ and then $I_{A:B}(\rho_{\tilde A B})=S(\rho_B)-\sum_{i=0}^{d-1}p_iS(\rho_B^i)$. The equivalence between Eqs (\ref{eq:qdi2}) and (\ref{eq:qdi3}) is obtained directly by definition.
\end{proof}
\end{theorem}

Theorem \ref{th:qdi} indicates that the quantum discord based on incoherent measurement equals to the basis-dependent discord defined in Ref. \cite{PhysRevX.6.041028}. As proved in Ref. \cite{PhysRevX.6.041028}, the basis-dependent discord vanishes not only for incoherent-quantum states, but also for coherent states which have a decomposition
\begin{equation}
\rho_{AB}=\sum_j\rho_A^j\otimes\rho_B^j,\label{eq:zero_QDI}
\end{equation}
such that all $\rho_A^j$ are perfectly distinguishable by the incoherent projective measurement. This result is natural by using Definition \ref{def:qdi}. The incoherent correlation on $A$ reaches the mutual information if there exist a local incoherent measurement on $A$ which can reveal the mutual information between $A$ and $B$. That is to say, the bipartite state is separable with each $\rho_A^j$ distinguishable by some incoherent measurement.

\section{Behavior of QDI under local incoherent operations}
Similar to the local creating property of the traditional discord, the discord based on incoherent measurement can also be created by local incoherent operations. In the following, we study the behavior of $D_{B|A}(\rho_{AB})$ under local operations.

(P1) $D_{B|A}(\rho_{AB})$ does not change under local unitary on $B$ or local incoherent unitary on $A$.
\begin{proof}
Let $\rho'_{AB}=U_A^\mathrm{I}\otimes U_B\rho_{AB}U_A^{\mathrm{I}\dagger}\otimes U_B^\dagger$, where $U_A^\mathrm{I}$ and $U_B$ are arbitrary incoherent unitary on $A$ and unitary on $B$. Because an incoherent unitary $U^\mathrm{I}$ satisfies the commutative property $U^\mathrm{I}\Delta(\cdot)U^{\mathrm{I}\dagger}=\Delta[U^\mathrm{I}(\cdot)U^{\mathrm{I}\dagger}]$, we have $\rho'_{\tilde AB}=\Delta_A\otimes\iden_B(\rho_{AB})=U_A^\mathrm{I}\otimes U_B\rho_{\tilde AB}U_A^{\mathrm{I}\dagger}\otimes U_B^\dagger$. Furthermore, local unitary does not change the mutual information, so $I(\rho'_{AB})=I(\rho_{AB})$ and $I(\rho'_{\tilde AB})=I(\rho_{\tilde AB})$. It follows from Eq. (\ref{eq:qdi2}) that $D_{B|A}(\rho'_{AB})=D_{B|A}(\rho_{AB})$.
\end{proof}

(P2) $D_{B|A}(\rho_{AB})$ can not be increased by local operations on $B$.
\begin{proof}
Here we first prove that discarding a subsystem on $B$ side does not increase the QDI defined on $A$, i.e., $D_{BB'|A}(\rho_{ABB'})\geq D_{B|A}(\rho_{AB})$, where $\rho_{AB}=\tr_{B'}(\rho_{ABB'})$. To this end, we employ Eq. (\ref{eq:qdi3}) and obtain
\begin{eqnarray}
& & D_{BB'|A}(\rho_{ABB'})- D_{B|A}(\rho_{AB})\nonumber\\
&=& [C_r(\rho_{ABB'})-C_r(\rho_{\tilde ABB'})-C_r(\rho_A)]\nonumber\\
&&-[C_r(\rho_{AB})-C_r(\rho_{\tilde AB})-C_r(\rho_A)] \nonumber\\
&=& [C_r(\rho_{ABB'})-C_r(\rho_{AB})-C_r(\rho_{B'})]\nonumber\\
&&-[C_r(\rho_{\tilde ABB'})-C_r(\rho_{\tilde AB})-C_r(\rho_{B'})]\nonumber\\
&=&I_{AB:B'}(\rho_{ABB'})-I_{AB:B'}(\rho_{\tilde ABB'})\geq0.
\end{eqnarray}
The last inequality is because local operations can not increase mutual entropy.

Any operation $\Lambda_B$ on $B$ can be realized by appending an ancilla $B'$, applying a local unitary $U_{BB'}$ on $B$ and $B'$, and then discarding $B'$. From Eq. (\ref{eq:qdi2}), appending an ancilla on $B$ side does not change the QDI on $A$. Hence we have
\begin{eqnarray}
D_{B|A}(\rho_{AB})&=&D_{BB'|A}\left(U_{BB'}(\rho_{AB}\otimes\rho_{B'}) U_{BB'}^\dagger\right)\nonumber\\
&\geq&D_{B|A}\left(\iden_A\otimes\Lambda_B(\rho_{AB})\right).
\end{eqnarray}
This completes the proof.
\end{proof}

This monotonic property makes $D_{B|A}(\rho_{AB})$ a proper quantification of correlations. Actually, the traditional quantum discord also satisfies this property.

Also, (P2) is a generalization of a result in Ref. \cite{PhysRevA.95.042328}, which says that the remaining coherence defined as $C^{\mathcal T}(\rho_{AB})=C_r(\rho_{AB})-C_r(\rho_{\tilde AB})-C_r(\rho_{A\tilde B})$ is nonnegative. From Eq. (\ref{eq:qdi3}), $C^{\mathcal T}(\rho_{AB})\geq0$ is equivalent to $D_{B|A}(\rho_{AB})\geq D_{B|A}(\rho_{A \tilde B})$, which is a spacial case of (P2). Ref. \cite{PhysRevA.95.042328} points out that $C^{\mathcal T}(\rho_{AB})=0$ for states with vanishing $D_{B|A}$ or $D_{A|B}$, but leaves it open whether other states satisfy this equation. Here we give a positive answer to this problem. To this end, we consider the maximal entangled state $\rho^\mathrm{m}_{AB}=|\Psi\rangle\langle\Psi|$ with $|\Psi\rangle=\frac{1}{\sqrt2}(|+0\rangle+|-1\rangle)$ and $|\pm\rangle=\frac{1}{\sqrt2}(|0\rangle\pm|1\rangle)$. Employing Eq. (\ref{eq:qdi2}), we obtain $D_{B|A}(\rho^\mathrm{m}_{AB})=D_{B|A}(\rho^\mathrm{m}_{A\tilde B})=1$. It means that the completely dephasing map on $B$ causes equal amount of decrease in the total correlation and the incoherent correlation.

Another related question is whether the monogamy relation holds when multipartite systems are considered, i.e., whether $D_{BB'|A}(\rho_{ABB'})$ is no less than $D_{B|A}(\rho_{AB})+D_{B'|A}(\rho_{AB'})$ for any tripartite state $\rho_{ABB'}$. We give a negative answer to this question. Actually, for GHZ state $|\mathrm{GHZ}\rangle=\frac{1}{\sqrt2}(|000\rangle+|111\rangle)$, the monogamy relation $D_{B|A}(\rho^\mathrm{GHZ}_{AB})+D_{B'|A}(\rho^\mathrm{GHZ}_{AB'})-D_{BB'|A}(\rho^\mathrm{GHZ}_{ABB'})=-1<0$ holds, but for $|\mathrm{W}\rangle=\frac{1}{\sqrt3}(|001\rangle+|010\rangle+|100\rangle)$, we have $D_{B|A}(\rho^\mathrm{W}_{AB})+D_{B'|A}(\rho^\mathrm{W}_{AB'})-D_{BB'|A}(\rho^\mathrm{W}_{ABB'})=2-\log_23>0$. Generally, $D_{B|A}(\rho_{AB})+D_{B'|A}(\rho_{AB'})-D_{BB'|A}(\rho_{ABB'})=I_{B:B'|\tilde A}(\rho_{ABB'})-I_{B:B'|A}(\rho_{ABB'})$, where $I_{A:B|C}(\rho_{ABC})=S(\rho_{AC})+S(\rho_{BC})-S(\rho_{ABC})-S(\rho_C)$ is the conditional mutual information. Our results show that $I_{A:B|C}(\rho_{ABC})$ is not monotonic under local operations on $C$.

(P3) $J_{B|A}(\rho_{AB})$ can not be increased by coherence non-activating operations. If a quantum operation $\Lambda$ satisfies the coherence non-activating condition $\Delta\circ\Lambda=\Delta\circ\Lambda\circ\Delta$, then $J_{B|A}(\rho_{AB})\geq J_{B|A}\left(\Lambda_A\otimes\iden_B(\rho_{AB})\right)$.
\begin{proof}
We first show that a measurement $M=\{M_\mu\}_\mu$ on party $A$ of $\rho'_{AB}=\Lambda\otimes\iden_B(\rho_{AB})$ is equivalent to the measurement $M'=\{\Lambda^*(M_\mu)\}_\mu$ on party $A$ of $\rho_{AB}$, where $\Lambda^*(\cdot)=\sum_lK_l^\dagger(\cdot)K_l$ and $K_l$ are the Kaus operator of $\Lambda$. Here $M'$ is a quantum measurement because its elements $\Lambda^*(M_\mu)$ are positive and satisfy $\sum_\mu \Lambda^*(M_\mu)=\iden$. The measurement $M$ on party $A$ of $\rho'_{AB}$ gives the probability
\begin{eqnarray}
p'(M_\mu)&=&\tr\left((M_\mu\otimes\iden_B)\Lambda\otimes\iden_B(\rho_{AB})\right)\nonumber\\
&=&\tr\left(\Lambda^*(M_\mu)\otimes\iden_B\rho_{AB}\right)=p(M'_\mu),
\end{eqnarray}
which is just the probability of the measurement $M'$ on party $A$ of $\rho_{AB}$. Similarly, the resulted states of $B$ read $\rho'_{B|M_\mu}=\tr_A\left((M_\mu\otimes\iden_B)\Lambda\otimes\iden_B(\rho_{AB})\right)=\tr_A\left(\Lambda^*(M_\mu)\otimes\iden_B\rho_{AB}\right)=\rho_{B|M'_\mu}$.

If $\Lambda$ is coherence non-activating, then $\Lambda^*$ preserves the incoherence of measurement, because $\Lambda^*\circ\Delta^*=\Delta^*\circ\Lambda^*\circ\Delta^*$ and $\Delta^*=\Delta$. From Theorem \ref{th:qdi}, the optimal measurement which reach the maximum in the definition of incoherent correlation is the projective measurement $\{|j\rangle\langle j|\}$. Hence, the incoherent correlation in $\rho'_{AB}$ reads
\begin{eqnarray}
J_{B|A}(\rho'_{AB})&=&S(\rho_B)-\sum_j p'_jS(\rho'_{B|j}),\nonumber\\
&=&S(\rho_B)-\sum_j p^{M'_j}S(\rho_{B|M'_j})
\end{eqnarray}
where $\rho_B=\tr_A(\rho'_{AB})=\tr_A(\rho_{AB})$, $M'_j=\Lambda^*(|j\rangle\langle j|)$ $p'_j=\tr(|j\rangle_A\langle j|\rho'_{AB})=\tr(\Lambda^*(|j\rangle_A\langle j|)\rho_{AB})=p^{M'_j}$, and $\rho'_{B|j}=\tr_A(|j\rangle_A\langle j|\rho'_{AB})=\tr_A(\Lambda^*(|j\rangle_A\langle j|)\rho_{AB})=\rho_{B|M'_j}$. Because $\{M'_j\}$ is an incoherent measurement, but may not be the one that reaches the maximum in the definition of $J_{B|A}(\rho_{AB})$, we have $J_{B|A}(\rho'_{AB})\leq J_{B|A}(\rho_{AB})$.
\end{proof}

(P4) $D_{B|A}(\rho_{AB})$ can be created by local incoherent operations on $A$.

In the spacial case where $A$ is a qubit, if a state has vanishing QDI, then it is either an incoherent-quantum state $\rho^{\mathrm{iq}}=p|0\rangle\langle 0|\otimes\rho_B^0+(1-p)|1\rangle\langle 1|\otimes\rho_B^1$ or a product state; hence, QDI can not be created by maximal incoherent operations (MIO). Generally, states with vanishing QDI is in the form of Eq. (\ref{eq:zero_QDI}), where $\rho_A^j$ can be coherent. In Ref. \cite{PhysRevX.6.041028}, an example is given to show that QDI can be created by local incoherent operations on $A$. Also, they prove that QDI is a monotone under genuine incoherent operations (GIO). Nevertheless, there are other coherence non-generating operations which can not create QDI. For example, the qutrit channel with Kraus operators as $K_0=\frac{1}{\sqrt2}(|-_{01}\rangle\langle0|+|1\rangle\langle1|)$, $K_1=\frac{1}{\sqrt2}(|+_{01}\rangle\langle0|+|0\rangle\langle1|)$, and $K_2=|2\rangle\langle2|$ is in MIO but not in IO (and hence not in GIO). This channel can not create QDI. Actually, it breaks the QDI in any qutrit-qudit states.


Here we observe that the ability of a quantum channel to create QDI can be activated by a parallel identical channel. Precisely, although a channel $\Lambda_A$ can not create QDI in any state $\rho_{AB}$ with vanishing $D_{B|A}$, it is possible that $\Lambda_A\otimes\iden_{A'}$ can create $D_{B|AA'}$. As an example, we consider the initial state $\rho_{AA'B}=\frac12 |000\rangle\langle000|+\frac14(|01\rangle+|10\rangle)(\langle01|+\langle10|)\otimes|1\rangle\langle1|$, which has zero QDI on $AA'$. Now let a depolarizing channel $\Lambda_2^{\mathrm{dep}}(\rho)=p\rho+(1-p)\frac{\iden}{2}$ acting on the qubit $A$, and the three-qubit state becomes $\rho'_{AA'B}=\frac12 \left[p|00\rangle\langle00|+(1-p)\frac{\iden}{2}\otimes|0\rangle\langle0|\right]\otimes|0\rangle\langle0|+\frac14\left[p(|01\rangle+|10\rangle)(\langle01|+\langle10|)+(1-p)\iden\otimes\iden\right]\otimes|1\rangle\langle1|$. It can be shown that $D_{B|AA'}(\rho'_{AA'B})>0$ for $0<p<1$. Therefore, although a qubit depolarizing channel $\Lambda^{\mathrm{dep}}_2$ (which is in MIO) can not create QDI, the tensor product channel $\Lambda^{\mathrm{dep}}_2\otimes \iden$ has the ability to create QDI.

Now we define the completely QDI non-generating channel as follows. A quantum channel $\Lambda$ is completely QDI non-generating if $\Lambda\otimes\iden$ does not has the ability to create QDI.

\begin{proposition}
A quantum channel is completely QDI non-generating if and only if it is a composition of GIO and incoherent unitary operations.
\begin{proof}
For the ``if'' part, because incoherent unitary does not change the QDI, we only need to prove that GIO are completely QDI non-generating. If $\Lambda\in\gio$, then we have $\Lambda\otimes\iden(|ij\rangle\langle ij|)=\Lambda(|i\rangle\langle i|)\otimes |j\rangle\langle j|=|ij\rangle\langle ij|$, which means that $\Lambda\otimes\iden$ is also in GIO. Therefore, $\Lambda\otimes\iden$ does not has the ability to create QDI, and then $\Lambda$ is completely QDI non-generating.

For the ``only if'' part, let us consider the following tripartite state $\rho_{AA'B}=\frac1d\sum_{j=0}^{d-1}|j\rangle_A\langle j|\otimes |\phi_j\rangle_{A'}\langle\phi_j|\otimes|j\rangle_B\langle j|$, where $\{|j\rangle_A\}_{j=0}^{d-1}$ is the incoherent basis of $A$, $|\phi_j\rangle_{A'}$ are linearly independent states of $A'$ which can not be distinguished by incoherent measurement. By definition, $D_{B|AA'}(\rho_{AA'B})=0$. If $\Lambda$ is completely QDI non-generating, then $\Lambda_A(\rho_{AA'B})=\frac1d\sum_{j=0}^{d-1}\Lambda(|j\rangle_A\langle j|)\otimes |\phi_j\rangle_{A'}\langle\phi_j|\otimes|j\rangle_B\langle j|$ has vanishing QDI on $AA'$. It means that $\Lambda(|j\rangle_A\langle j|)\otimes |\phi_j\rangle_{A'}\langle\phi_j|$ can be perfectly distinguished from each other by incoherent measurement. Because these states are product states and $|\phi_j\rangle_{A'}$ are indistinguishable by incoherent measurement, the $d$ states $\Lambda(|j\rangle_A\langle j|)$ can be distinguished by incoherent measurement. It follows that $\{\Lambda(|j\rangle_A\langle j|)\}$ is also the incoherent basis, i.e. there exist an incoherent unitary $U_I$ such that $\Lambda(|j\rangle_A\langle j|)=U_I|j\rangle\langle j|U_I^\dagger,\ \forall j$. Hence, $U_I^\dagger\Lambda(\cdot)U_I$ is GIO, and $\Lambda$ is a composition of GIO and incoherent unitary operations.
\end{proof}
\end{proposition}

\section{Conclusion}
The coherence non-activating measurement, as well as the quantum discord based on it, have been explicitly studied. If a POVM gives the same result when the coherence in a quantum state is destroyed, then its measurement elements should be diagonal in the incoherent basis. In order to witness that a POVM is not an incoherent measurement, we derive an inequality (Eq. (\ref{eq:im_steering})), and show that it is tight when projective measurement with white noise is considered.

When the set of POVM in the definition of classical correlation and quantum discord is restricted to the coherence non-activating measurement, we obtain the incoherent correlation and the quantum discord based on incoherent correlation (QDI). Different from the traditional quantum discord, where the optimal POVM may not be a projection, the optimal incoherent measurement in the definition of QDI is always the projection to the incoherent basis.

The incoherent correlation and QDI defined on $A$ do not change under unitary on $B$ or incoherent unitary on $A$, and do not increase under any local operations on $B$. The monogamy relation for QDI does not hold in general. The incoherent correlation is monotonically decreasing under coherence non-activating quantum operations. QDI defined on $A$ can be created by local coherence non-generating operations on $A$. Although the set of QDI non-generating operations is not subset to IO, the ability of some operations to create QDI can be activated by a parallel identity operation. We define the completely QDI non-generating channels as the quantum operations which can not create QDI even if a parallel identity operation is employed, and prove that a quantum channel is completely QDI non-generating if and only if it is a composition of GIO and incoherent unitary operations.

\begin{acknowledgments}
This work was supported by NSFC under Grant No. 11774205, and Young Scholars Program of Shandong University.
\end{acknowledgments}

\bibliography{apssamp}

\end{document}